\documentclass[vecphys]{svmult}

\begin{document}

\def\simlt{\lower.5ex\hbox{$\; \buildrel < \over \sim \;$}}
\def\simgt{\lower.5ex\hbox{$\; \buildrel > \over \sim \;$}}
\def\simpropto{\lower.2ex\hbox{$\; \buildrel \propto \over \sim \;$}}

\title*{Galaxy Formation and  Dark Matter }
\author{Joseph Silk}
\institute{Department of Physics, Denys Wilkinson Building,
Keble Road,  Oxford, OX1 3RH, UK\\
\texttt{silk@astro.ox.ac.uk}}
%
%

\maketitle\begin{abstract}
The challenge of dark matter may be addressed 
in two ways; by studying the confrontation of 
structure formation with observation and by direct and indirect searches.
In this review, I will focus on those aspects of dark matter
that are relevant for understanding galaxy formation, and describe the
outlook for detecting the most elusive  component, non-baryonic dark 
matter. Galaxy formation theory is driven by
phenomenology and by numerical simulations of dark matter clustering
under gravity.  Once the complications of star formation are
incorporated, the theory becomes so complex that the brute force
approach of numerical simulations needs to be supplemented by incorporation of  such astrophysical processes as feedback by supernovae and by active galactic nuclei.  I present 
a few semi-analytical perspectives that may shed some insight into
the nature of galaxy formation.
 
\end{abstract}

\section{Introduction}

Dark matter dominates over ordinary matter. The observations are
compelling. Of course, by definition we do not observe matter if it is
dark.  Minimal gravitational  theory is needed to take us from the observational
plane to conclude that dark matter is required.  Gravity has been tested 
over scales that range from
millimetres to megaparsecs. Newton's description of gravity is
perfectly adequate, apart from generally small deviations due to the
curvature of space near massive objects, such as stars, or more
radically, black holes. Einstein's theory of gravity tells us that
gravity curves space and measuring this effect was one of the great
triumphs of 20th century physics. Nevertheless, pending its direct
detection, dark matter remains a hypothesis that depends, inevitably,
on our having the correct theory of gravitation. For the remainder of this review,
however, I will assume the reality  of dark matter dominance on scales from 
galactic to those spanning the entire universe.

The standard (or concordance) model of cosmology has a predominance of
dark energy. which amounts to
$65\%$ of the
mass energy today whereas non-baryonic matter is $30\%.$ In contrast,
luminous baryons (mostly in stars) constitute $0.5\%$
towards the total.
An important component of the standard model is the spectrum of
primordial density fluctuations, measured in the linear regime via the
temperature anisotropies of the CMB. This provides the initial
conditions for large-scale structure and galaxy formation via
gravitational instability once the universe is matter-dominated. Dark
matter consequently provides the gravitational potential wells within which
galaxies formed. The dark matter and galaxy formation paradigms are
inextricably interdependent. Unfortunately we have not yet identified
a dark matter candidate, nor do we yet understand the fundamental aspects
of galaxy formation. Nevertheless, cosmologists have not been
deterred, and have even been encouraged to develop novel probes and
theories that seek to advance our understanding of these forefront
issues.

Progress has been made on the baryonic dark matter front. Only about
half of the baryons initially present in galaxies, or more precisely,
on the comoving scales over which galaxies formed, are directly
observed. We cannot predict with any certainty the mass fraction in
dark baryons. Yet there are excellent candidates for the dark baryons,
both compact and especially diffuse.

In contrast, we have at least one elegant and moderately compelling
theory of particle physics, SUSY, that predicts the observed fraction
of nonbaryonic dark matter. Unfortunately, we have no idea yet as to
whether the required stable supersymmetric particles actually exist.

In this review, I will first describe the increasingly standard
precision model of cosmology that enables us to provide an
inventory of cosmic baryons.  I summarise the current situation with regard to possible baryonic dark matter.  I discuss how nonbaryonic matter has
been successfully used to provide an infrastructure for galaxy
formation, and review the astrophysical issues, primarily centering on
star formation and feedback.  I conclude with  the outlook for future
progress.
 for nonbaryonic dark matter detection and galaxy formation.

\section{Precision cosmology}

Modern cosmology has emphatically laid down a challenge to
theorists. A combination of new experiments has unambiguously measured
the key parameters of our cosmological model that describes the
universe.  These include the temperature fluctuations in the cosmic
microwave background, the large galaxy redshift surveys, gravitational shear distortions of distant galaxies by lensing, the studies
of the intergalactic medium via the distribution of absorbing neutral
clouds along different lines of sight and the use of distant Type Ia
supernovae as standard candles.  Cosmologists now
debate the error bars of the standard model parameters. The
ingredients of the standard model in effect define the model.  These
most crucially are the Friedmann-Robertson-Walker metric and the Friedmann-Lemaitre
equations, and the contents of the universe: baryons, neutrinos,
photons, baryons, dark matter and dark energy.  On these
constituents is superimposed a distribution of primordial adiabatic
density (scalar) fluctuations characterised by a power spectrum of
specified amplitude and spectral index.  In addition, there may be a
primordial gravity wave tensor mode of fluctuations.  The number of
free parameters in the standard model is 14, of which the most
significant are: $H_0,\,\Omega_b,\,\Omega_m,\,\Omega_\Lambda,\,
\Omega_\gamma,\,\Omega_\nu,\,\sigma_8,\,n_s,\, r,\, n_T,\, $ and
$\tau.$ One can also add an equation of state for dark energy
parameter, $w=-p_\Lambda/\rho_\Lambda, $ in effect really a function
of redshift, and a rolling scalar (and possibly tensor) index,
$dn_s/dlnk.$

No single observational set constrains all, or even most, of these
parameters.  There are well-known degeneracies, most notably between
$\Omega_\Lambda$ and $\Omega_m,$ $\sigma_8$ and $\tau,$ and $\sigma_8$
and $\Omega_m.$ However use of multiple data sets helps to break these
degeneracies. For example, CMB anisotropies fix the combination
$\Omega_m + \Omega_\Lambda$ if a Hubble constant prior is adopted, 
as well as
 $\Omega_b h^2$ 
and $\Omega_m h^2,$
and SNIa constrain the (approximate) combination
$\Omega_m - \Omega_\Lambda.$ Both weak lensing and peculiar velocity
surveys specify the product $\Omega_m^{-0.6} \sigma_8.$ Lyman alpha
forest surveys extend the latter measurement to Mpc comoving scales,
probing the currently nonlinear regime.  Finally, baryon oscillations
are providing a measure of $\Omega_m/\Omega_b,$ independently of the
CMB. Interpretation in terms of a standard model (Friedmann-Lemaitre
plus adiabatic fluctuations) yields the concordance model with
remarkably small error bars \cite{sel05}.

The flatness of space is measured to be
$\Omega_{total}=1.02\pm 0.02.$ Dark energy in the form of a
cosmological constant dominates the universe, with
$\Omega_{\Lambda}=0.72 \pm 0.02.$ The dark energy equation of state is
indistinguishable from that of a cosmological constant, with $w\equiv
p_\Lambda/\rho_\Lambda c^2=-0.99 \pm 0.1,$ this uncertainty holding to
$z\sim 0.5.$ Even at $z\sim 1,$ the claimed uncertainty around $w=-1$
is only 20 percent. Non-baryonic dark matter dominates over baryons
with $\Omega_m = 0.27\pm 0.02$ and $\Omega_b = 0.044\pm 0.004.$ Most
of the baryons are non-luminous, since $\Omega_\ast \approx 0.005.$

The spectrum of primordial density fluctuations is unambiguously
measured both in the CMB and in the large-scale galaxy distribution
from deep redshift surveys, and found to be approximately
scale-invariant, with scalar index $n_s=0.98\pm 0.02.$ One can also
constrain a possible relic gravitational wave background, a key
prediction of inflationary cosmology, by the tensor mode limit on relic
gravitational waves: $T/S<0.36.$ It has been argued that a fundamental
test of inflation requires sensitivity at a level $T/S \simgt 0.01$ \cite{boy05} .
Neutrinos are known to have mass as a consequence of atmospheric
($\nu_\tau,\nu_\mu$) and solar ($\nu_\mu, \nu_e$) oscillations, with a
deduced mass in excess of 0.001 eV for the lightest neutrino. From the
power spectrum of the density fluctuations, the inferred mass limit
(on the sum of the 3 neutrino masses) is $\Sigma m_\nu< 0.4 \rm eV.$

However one note of caution should be added. These tight error bars
all depend on adoption of  simple priors. If these are extended, to allow,
for example, for an admixture of generic primordial isocurvature
fluctuations, the error bars on many of these parameters increase
dramatically, by up
to an order of magnitude.

Clearly, the devil is in the observational details. Popular models of
inflation predict that $n\approx 0.97.$ Space is expected to be very close to
 flat, with $\Omega=1+\cal{O}$${(10^{-5})}.$  The numbers of
rare massive objects at high redshift is specified by the theory of gaussian random fields applied to the  primordial linear density fluctuations. The universe as viewed in the CMB
should be isotropic.  Any deviations from these predictions would be
immensely exciting.

Suppose deviations were to be found. This would allow all sorts of
possible extensions to the standard model of cosmology. One might
consider the signatures of string relics of superstrings or
transplanckian features in $\delta T/T|_k$ \cite{gasp05}.
 Large-scale cosmology
might be affected by compact topology or global anisotropy with observable 
signatures in CMB temperature and polarisation maps \cite{riaz06}. The
initial conditions might involve primordial nongaussianity. 
Anthropically constrained landscape scenarios of the metauniverse 
prefer a slightly open universe \cite{suss05}.
Some of
these features, and others, could be a consequence of compactification
from higher dimensions.

\section{The global baryon inventory}

There are several independent approaches to obtaining the baryon
 abundance in the universe. At $z\sim 10^9,$ primordial
 nucleosynthesis of the light elements yields $\Omega_b=0.04\pm
 0.004.$ At the epoch of matter-radiation decoupling, $z\sim 1000,$
 the ratios of odd and even CMB acoustic peak heights set $\Omega_b
 =0.044\pm 0.003.$ At more recent epochs, Lyman alpha forest modelling
 of the intergalactic medium at $z\sim 3$ as viewed in absorption
 along different lines of sight towards high redshift quasars at
 $z\sim 3$ yields $\Omega_b\approx 0.04.$ At the present epoch, on
 very large scales, of order 10 Mpc comoving linear regime equivalent,
 the intracluster baryon fraction measured via x-ray observations of
 massive galaxy clusters provides a baryon fraction of $15\% .$ This
 translates into $ \Omega_b\approx 0.04.$ In summary, we infer that
 $\Omega_b=0.04 \pm 0.005$ and $\Omega_b/\Omega_m =0.15\pm 0.02.$

One's immediate impression is that, at least until very recently,  most of the baryons in the universe
today are not accounted for. The reasoning is as follows. The luminous
content in the form of stars sums to $\Omega_b\approx 0.004$ or $10\%$
in spheroids, and $\Omega_b\approx 0.002$ or $5\%$ in disks.  There is
also hot intracluster gas amounting to $\Omega_b\approx 0.002$ or
$5\%$.  Current epoch observations of the cold/warm photo-ionised IGM
via the nearby Lyman alpha/beta forest at $10^4-10^5$K as well as CIII
(at $z\sim 0$) yield a much larger baryonic reservoir of gas,
$\Omega_b\approx 0.012$ or $30\%$. This gas is metal-poor, with an
abundance of about 10\% solar \cite{dan05}.
So far, we have only accounted  for $50\%$ of current epoch baryons.

The probable breakthrough, however, has come with  recent detections
of the warm-hot intergalactic medium at $T\simlt 10^5-10^6\rm K$ at
$z\sim 0,$ observed in OVI absorption in the UV and especially via
x-ray absorption via OVII and OVIII hydrogen-like transitions towards
low redshift luminous AGN. Something like $\Omega_b\approx 0.012$ or
$30\%$ of the primordial baryon fraction appears to be in this form,
enriched (in oxygen, at least) to about 10\% of the solar value
\cite{nic05}. We now have $\simgt 80\%$ of the baryons accounted for
today.  The total baryon content sums to $\Omega_b=0.032 \pm 0.005.$
Given the measurement uncertainties, this would seem to remove any
strong case for more exotic forms of dark baryons being present.

However, the situation is not so simple. The Andromeda Galaxy and our
 own galaxy are especially well-studied regions, where dark matter and
 baryons can be probed in detail. In the Milky Way Galaxy, the virial
 mass out to 100 kpc is $M_{virial}\approx 10^{12}\rm M_\odot,$
 whereas the baryonic mass, mostly in stars, is $M_{\ast} \approx
 6-8\times 10^{10}\rm M_\odot. $ The inferred baryon fraction is at
 most $8\% $ \cite{klyp02}. Similar statements may be made for massive 
elliptical galaxies
\cite{lint05}. These  in fact are upper limits as the dark mass estimate
 is a lower bound. 

I infer that globally, there is no problem.  Nevertheless the
outstanding question is:  where are the galactic baryons? Most of the
baryons are globally accounted for. But this is not the case for our
own galaxy and most likely for all comparable galaxies.  We cannot
account for a mass in baryons comparable to that in stars.  It is
possible that up to $ 10\%$ of all the baryons {\it may} be dark, and
that the dark baryons are comparable in mass to the galactic stars.

\section{The ``missing" baryons}
There are several possibilities for the ``missing" baryons. Perhaps
they never were present in the protogalaxy. Or they are in the outer
galaxy. Or, finally, they may have been ejected.

The first of these options seems very unlikely 
(although we return below to a variant on this).
Consider the second option. The most likely candidates for dark
baryons are massive baryonic objects or MACHOs.  These are constrained
by several gravitational microlensing experiments.  The allowed mass
range is between $10^{-8}$ and $10\,\rm M_\odot ,$ and the best
current limit on the MACHO abundance is $\simlt$ 20\% of the dark halo
mass.  In fact, one experiment, that of the MACHO Collaboration,
claims a detection from some 20 events seen towards the LMC, most of
which cannot be accounted for by star-star microlensing. The observed
range of amplification time-scales specifies the mass of the lensing
objects. The preferred MACHO mass is around $\sim 0.5\,\rm M_\odot.$

This mass favours an interpretation in terms of old halo white
dwarfs. Main sequence stars in this mass range can be excluded.
Current searches for halo high velocity old white dwarfs utilise the
predicted colours and proper motions as a discriminant from field
dwarfs, and set a limit of $\simlt 4\%$ of the dark halo mass on a
possible old white dwarf component in the halo \cite{cre05}. However
even if this limit were to apply, an extreme star formation history
and protogalactic IMF would be required. Observations at high redshift
both of star-forming galaxies and of the diffuse extragalactic light
background, combined with chemical evolution and SNIa constraints,
make such an hypothesis extremely implausible.

If the empirical mass range constraint is relaxed, theory does not
exclude either primordial brown dwarfs ($0.01-0.1\,\rm M_\odot$),
primordial black holes (mass $\simgt 10^{-16}\rm M_\odot$) or even
cold dense $H_2$ clumps $\simlt 1 \,\rm M_\odot.$ The latter have been
invoked in the Milky Way halo in order to account 
for extreme halo scattering events  \cite{wal98} or unidentified submillimetre 
 sources \cite{law01}.  However these
possibilities seem to be truly acts of the last resort in the absence
of any more physical explanation.

There is indeed another possibility that seems far less ad hoc.  The
nearby intergalactic medium is enriched to about 10\% of the solar
metallicity, and contains of order 50\% of the baryons in
photo-ionised and collisionally ionised phases.  This strongly
suggests that ejection from galaxies via early winds must have
occurred, and moreover would inevitably have expelled a substantial
fraction of the baryons along with the heavy elements. Supporting
evidence comes from x-ray observations of nearby galaxy groups, which
demonstrate that many of these are baryonically closed systems,
containing their prescribed allotment of baryons.

There are candidates for young galaxies undergoing extensive mass loss
 via winds. These are the Lyman break galaxies at $z\sim 2-4.$
 Observations of spectral line displacements of the interstellar gas
 relative to the stellar component as well as of line widths are
 indicative of early winds from $L_\ast$ galaxies \cite{adel03}. 
Studies of nearby
 starburst galaxies, essentially lower luminosity counterparts of the
 distant LBGs, show that the gas outflow rate in winds is of order the
 star formation rate. The intracluster medium to $z\sim 1$ is enriched
 to about a third of the solar metallicity, again suggestive of
 massive early winds, in this case from early-type galaxies.  Hence
 the ``missing'' baryons could be in the IGM, with about as much mass
 ejected in baryons as in stars remaining.

The ejection hypothesis however has to confront a theoretical
 difficulty. Winds from $\rm L_\ast$ galaxies cannot be reproduced by
 hydrodynamical simulations of forming galaxies \cite{spr03}. The momentum source
 for gas expulsion appeals to supernovae.  SN feedback works for dwarf
 galaxies and can explain the observed outflows in these
 systems. However an alternative feedback source is needed for massive
 galaxies. This most likely is associated with AGN, and the ubiquitous
 presence of central supermassive black holes in galaxy spheroids.

First, however, I address a more  pressing and not unrelated problem, namely 
given that 90 percent of the matter in the universe is nonbaryonic and cold, how well does CDM fare in confronting galaxy formation models?

\section{ Large-scale structure and cold dark matter: the issues}

The cold dark matter hypothesis has had some  remarkable successes in
confronting observations of the large-scale structure of the universe.
These have stemmed from predictions, now verified,  of the amplitude of the
temperature fluctuations in the cosmic microwave background that are
directly associated with the seeds of structure formation.  The
initial conditions for gravitational instability to operate in the
expanding universe were measured.  The formation of galaxies and
galaxy clusters was explained, as was the filamentary nature of the
large-scale structure of the galaxy distribution.  Nor was only the
amplitude confirmed as a prerequisite for structure formation. The
Harrison-Zeldovich-Peebles ansatz of an initially scale-invariant
fluctuation spectrum, later motivated by inflationary cosmology, has
now been confirmed over scales from 0.1 to 10000 Mpc, via a combination of
CMB, large-scale galaxy distribution and IGM measurements.

Despite these stunning successes, difficulties remain in reconciling
theory with observations. These centre on two aspects: the
uncertainties in star formation physics that render any definitive
predictions of observed galaxy properties unreliable, and the detailed
nature of the dark matter distribution on small scales, where the
simulations are also incomplete.

The former issues include such observables as the galaxy luminosity
function, disk sizes and mass-to-light ratios, and the presence of
old, red massive galaxies at high redshift. These difficulties in the
confrontation of galaxy formation theory and observational data are
plausibly resolved by improving the prescriptions for star formation
and feedback, although there are as yet no definitive answers. The
latter issues require high resolution dark matter simulations combined
with hydrodynamic simulations of the baryons including star formation
and feedback.

I will focus first on the dark matter conundrums, and in particular on
the challenges posed by theoretical predictions of dark matter
clumpiness, cuspiness and concentration. Implementation of numerical
simulations of dark halos of galaxies in the context of hierarchical
galaxy formation yields repeatable and reliable results at resolutions
of up to $\sim 10^5\,\rm M_\odot$ in $M_\ast$ halos.  It is clear that
the simulations predict an order of magnitude or more dwarf galaxy
halos than are observed as dwarf galaxies. It is more controversial but
probably true that the dark halos of dwarf galaxies and of barred
galaxies do not have the $\sim r^{-1}$ central cusps predicted by high
resolution simulations. The dark matter concentration parameter,
defined by the ratio of $r_{200},$ approximately the virial scale, to
the scale length, within which the cusp profile is found, measures the
cosmological density at virialisation, and hence should be
substantially lower for late-forming galaxy clusters than for galaxies. This may
not  be the case in the best-studied examples of massive
gravitationally lensed clusters, cf. \cite{ogu05}. There are also
examples of early-forming massive clusters \cite{over06}

\section{Resurrection via astrophysics}

There are at least two viewpoints about resolving the dark matter
issues, involving either fundamental physics or  astrophysics. 
Tinkering with fundamental physics, in essence, opens up a
Pandora's box of phenomenology. It seems to me that one should first
take the more conservative approach of examining the impact of
astrophysics on the dark matter distribution before advocating
more fundamental changes. Of course if one could learn about
fundamental physics, such as a new theory of gravity or higher
dimensional dark matter relics from dark matter modelling, this would
represent an unprecedented and unique breakthrough. But 
the prospect of such revelations  may be premature.

Astrophysical resolution involves two complementary approaches.
One  incorporates star and AGN
feedback in the dense baryonic core that forms by gas
dissipation. Massive gas outflows can effectively weaken the dark
matter gravity, at least in the central cusp.  These may include
stellar feedback driving  massive winds via supernovae
augmented by a  top-heavy IMF and/or by 
hypernovae, or the impact of  supermassive black hole-driven 
outflows. Another
mechanism that shows some promise in terms of generating an isothermal
dark matter core is dynamical feedback, via a central massive rotating
gas bar. Such bars may form generically and dissolve rapidly, but
their dynamical impact on the dark matter has not yet been fully evaluated
\cite{wei02, val03, ath04}.

All of these are radical procedures, but some are more radical than
others.  
To proceed, one has to better understand when and how galaxies formed.
Fundamental questions in galaxy formation theory still  remain unresolved.
Why do massive galaxies assemble early? And how can their stars form
rapidly, as inferred from the $\alpha/Fe$ abundance ratios?  Where are
the baryons today? And if, as observations suggest, they are in the
intergalactic medium, including both the photo-ionised Lyman $\alpha$ forest and
the collisionally ionised warm-hot intergalactic medium (WHIM), how
and when is the intergalactic medium (IGM) enriched to 0.1 of the
solar value?  Can the galaxy luminosity function be reconciled with
the dark matter halo mass function?  Does the predicted dark matter
concentration allow a simultaneous explanation of both the
Tully-Fisher relation, the fundamental plane  and the galaxy luminosity function?  And for that
matter, is the dark matter distribution consistent with barred galaxy and low
surface brightness dwarf galaxy rotation curves?

The observational data that motivates many of these questions can be
traced back to the colour constraints on the interpretation of galaxy
spectral energy distributions by population synthesis modelling 
\cite{san86,lar86}.
The galaxy distribution is bimodal in colour, and this can be seen
very clearly in studying galaxy clusters.  The presence of a red
envelope in distant clusters of galaxies testifies to the early
formation of massive ellipticals.  A major recent breakthrough has
been the realisation from UV observations with GALEX that many
ellipticals, despite being red, have an ongoing trickle of star
formation.  Most field galaxies and those on the outskirts of clusters
are blue, and are actively forming stars.

The general conclusion is that there must be two modes of global star
formation: quiescent and starburst. The inefficient, long-lived,
disk mode is motivated by cold gas accretion and global disk
instability. The low efficiency is due to negative feedback.  The disk
mode is relatively quiescent and continues to form stars for a Hubble
time. The violent starburst mode is necessarily efficient
as inferred from the $[\alpha/Fe]$ clock.  It is motivated by
mergers, including observations and simulations, as well as by CDM
theory.  The high efficiency is presumably due to positive feedback, but 
it is not clear how the feedback is provided.

\section{What determines the mass of a galaxy?}

The luminosity function of galaxies describes the stellar mass
function of galaxies. It is biased by star formation in the B (blue)
band but is a good tracer in the near-infrared (K) band. It is
sensitive to the halo mass, at least for spiral galaxies, as
demonstrated by rotation curves.  There is a characteristic
luminosity, and hence a characteristic stellar mass, associated with
galaxies: $L_\ast \approx 3\times 10^{10}\rm L_\odot$ and $M_\ast
\approx 10^{11}\rm M_\odot$. The luminosity function declines
exponentially at $L>L_\ast.$ This is most likely a manifestation of
strong feedback.

Consider first the mass-scale of a galaxy. There is no difference in
dark matter properties between galaxy, group or cluster scales, but
there is a very distinct difference in baryonic
appearance. Specifically, the baryons are mostly in stars below
a galaxy mass scale of $M_\ast$ and mostly in hot gas for systems much more 
massive than
 $M_\ast,$ such as galaxy groups \cite{mat05} and clusters.   A simple explanation
comes from considerations of gas cooling and star formation
efficiency. It does not matter whether the 
gas infall initially is  cold or whether it virialises during infall.
The gas generically will be clumpy, and cloud collisions will be at the virial velocity.
In order for the gas to form stars efficiently, a necessary condition 
is that the cooling time of the shocked gas be less than a dynamical time,
or 
$t_{cool} \simlt t_{dyn}.$

The inferred upper limit on the stellar mass, for stars to form within
a dynamical time in a halo of baryon fraction $f_b$ and mean density $\rho_h,$ 
can be written as
$$M_\ast=A^\beta m_p^{2\beta}G^{-(3+\beta)/2}(t_{cool}/t_{dyn})^\beta
 f_b^{1-\beta}\rho_h^{(\beta -1)/2},$$ where the cooling rate has been
 taken to be $\Lambda=Av_s^{2-3/\beta}$, with $\beta\approx 1$ being
 appropriate for metal-free cooling in the temperature range
 $10^5-10^6$K. This yields a characteristic mass $M_\ast/m_p \approx
 0.1\alpha^3\alpha_g^{-2}(m_p/m_e)(t_{cool}/t_{dyn})\approx 10^{68},$
 where $\alpha_g=Gm_p^2/e^2.$ This is comparable to the stellar mass
 associated with the characteristic scale in the Schechter fit to the
 luminosity function, and also the scale at which galaxy scaling
 relations change slope. However there is no reason to believe that
 the dynamical time argument gives as sharp a feature as is observed
 in the decline of the galaxy luminosity function to high
 luminosities.  Additional physics is needed.

\section{Outflows from disks}
In the quiescent mode, the clumpy nature of accretion suggests that
ministarbursts might occur.  In fact, what is more pertinent is the
runaway nature of supernova feedback in a cold gas-rich
disk. Initially, exploding stars compress cold gas and stimulate more
star formation.  Negative feedback is eventually guaranteed in part as
the cold gas supply is exhausted and also as the cold gas is ejected
in plumes and fountains from the disk, subsequently to cool and fall
back.

Global simulations have inadequate dynamical range to follow the
multiphase interstellar medium, supernova heating and star
formation. The following toy model provides an analytical description
of disk star formation.  I assume that self-regulation applies to the
hot gas filling factor $1-e^{-Q},$ where $Q$ is the porosity and is
defined by
\begin{eqnarray*}
 \left( SN \, bubble \, rate\right)\times \left(maximum \,\, bubble \,
\mbox{4-}volume \right) \\ \propto \left( star \ formation \
rate\right)\times \left( {turbulent\ pressure}^{-1.4}\right).
\end{eqnarray*}
One can now write the star formation rate as \cite{sil03}
$$\alpha_S \times rotation \, rate\times \, gas\, density $$ with $\alpha_S\equiv
Q\times\epsilon.$ Here $\epsilon= ({\sigma_{gas}}/{\sigma_{f}})^{2.7}$, where
the fiducial velocity dispersion $\sigma_f\approx 20\, {\rm km
s}^{-1}\left(E_{SN}/10^{51}\rm ergs\right)^{0.6}
\left(200M_\odot/m_{SN}\right)^{0.4}.$ 
Here $m_{SN}$ is the mass in stars formed per supernova and $E_{SN}$ is the 
initial kinetic energy in the supernova explosion.
The star formation efficiency
$Q\epsilon$ is
$$ 0.02 \left(\frac{\sigma_{gas}}{10 \, {\rm km s^{-1}}}\right)
\left(\frac{v_c}{400 \, {\rm km s^{-1}}}\right) \left(\frac{m_{SN}}{200 {\rm
M_\odot}}\right) \left(\frac{10^{51}{\rm ergs}}{E_{SN}}\right).$$ The
observed mean value is 0.017 \cite{ken98}.  Also, the analytic expression derived
for the star formation rate agrees with that found in 3-D multiphase
simulations \cite{sly05}.  In fact, the observed distribution of young stars
in merging galaxies cannot be fit by modelling the star formation rate with a
Schmidt-Kennicutt law, but requires the incorporation of a turbulence-like
term \cite{bar04}, as incorporated in this simple model.

To extract the wind, one might expect that the outflow rate equals the
product of the star formation rate, the hot gas volume filling factor,
and the mass loading factor ($f_L$).  This reduces to $ \sim Q^2 \epsilon\dot
M_\ast,$ or $\dot M_{outflow} \approx
f_L\alpha_S^2\epsilon^{-1}M_{gas}\Omega .  $ If $Q$ is of order 50\%,
then the outflow rate is of order the star formation rate, but this
evidently only is the case for dwarf galaxies. Once $\epsilon\gg 1$,  the
wind is suppressed.

This begs the question of how massive disks such as our own  and M31 have depleted their initial baryon content by of order 50 percent.
One cannot appeal to protospheroid outflows initiated by AGN (see below)
to resolve this issue.
Presumably baryon depletion in late-type massive disks
(with small spheroids) must have occurred  during the disk assembly phase. A collection of gas-rich dwarfs most likely assembled into a current epoch massive disk, and outflows from the dwarfs could plausibly have expelled of order half of the baryons into the Local Group or even beyond. However  weak lensing 
studies find that the typical late-type 
galaxy
in a cluster environment appears to have utilised its full complement of baryons over a Hubble time \cite{hoek05}, whereas an early-type galaxy may indeed have expelled about half of its baryons into the intracluster medium.

\section{Outflows from protospheroids}

Galaxy spheroids formed early. The inferred high efficiency of star
formation on a short time-scale, as inferred from the $\alpha/Fe $
enhancement, is suggestive of a feedback mechanism distinct from, and
much more efficient than, supernovae.

The preferred context for such a mechanism is that of ultraluminous
starbursts. Major mergers between galaxies produce extreme gas
concentrations that provide an environment for the formation of
supermassive black holes.  The observed correlation between SMBH mass
and the spheroid velocity dispersion suggests contemporaneous SMBH
growth and coupled formation of the oldest galactic stars.
 The spheroid stars are old and formed
when the galaxy formed.  Hence the SMBHs, which account via the
empirical correlation for approximately 0.001 of the spheroid mass,
must have formed in the protogalaxy more or less contemporaneously
with the spheroid.  Supermassive black hole growth is certainly
favoured in the gas-rich protogalactic environment.

Another clue is that both SMBHs, as viewed in AGN and quasars, and
massive galaxy spheroids formed anti-hierarchically at a similar
epoch, peaking at $z\sim 2.$ Massive systems form before less massive
systems. This could be a consequence of the same feedback mechanism,
which necessarily must be positive in order to favour the massive
systems. Supernova feedback is negative and is most effective in low
mass systems. SMBH outflows provide  an intriguing possibility 
for positive feedback that merits
further exploration. What is lacking for the moment is quantitative
evidence for the frequency with which AGN activity is associated with
ultraluminous infrared galaxies.  Nevertheless, AGN feedback seems to provide the most
promising direction for progress.

A specific mechanism  for positive feedback appeals to SMBH-induced outflows
interacting with the clumpy protogalactic medium.
Twin 
jets are accelerated from the vicinity of the SMBH along the minor
axis of the accretion disk. These jets are the fundamental power
source for the high non-thermal luminosities and the huge turbulent
velocities measured in the nuclear emission line regions in active
galactic nuclei and quasars.  The jets drive hot spots at a velocity
of order $0.1$c that impact the protogalactic gas. In a cloudy medium, 
the jets are frustrated and generate turbulence.
The jets are surrounded by hot cocoons that engulf and
overpressure ambient protogalactic clouds \cite{sax05}.  These clouds
collapse and form stars. The speed of the cocoon as it overtakes the
ambient gas clouds greatly exceeds the local gravitational
velocity. In this way, a coherent and positive feedback is provided
via triggering of massive star formation and supernovae on a 
time-scale shorter than the gravitational crossing time \cite{sil05} .
The short duty cycle for the AGN phase relative to the longer duty cycle for
the induced starburst must be incorporated into inferences from surveys
about the frequency  of associated AGN activity, if any.

Eventually, the input of energy must be highly disruptive for the
protogalaxy.  When the SMBH is sufficiently massive, its Eddington-limited
outflow drives out the remaining protogalactic gas in a wind. This curtailing
of spheroid growth allows one to understand the quantitative correlation
between SMBH mass and the spheroid gravitational potential \cite{sil98}.
Such negative feedback has been extensively applied in semi-analytic galaxy
formation simulations to stop the gas cooling that otherwise results in
excessive star formation in massive galaxies \cite{crot05}. 
However the possibility of positive feedback has not hitherto
been implemented.

\section{ULIGs and spheroid formation}

One may actually be seeing the AGN-triggering phenomenon at work in
 ultraluminous infrared galaxies (ULIGs),
 which
 plausibly are the sites of spheroid formation and SMBH growth, as well as in 
powerful radio galaxies. 
High velocity neutral winds are found both in NaI \cite{mart05} and in HI
 absorption \cite{morg05}
 against the central bright nuclei. 
The rate of mass ejected in these superwinds is
 inferred  to be a significant fraction of  the star formation rate. 
Hence the baryon mass
 ejected is likely to be of order the stellar mass formed. This helps account for the
 baryon budget, with a complementary mechanism involving supernovae operative
 in dwarf galaxies and the precursor phase of massive disks.

A simple analytic model of this phenomenon may be constructed as follows.
AGN momentum-driven outflow is inevitable once the mechanical momentum
luminosity $\dot M_wv_w$ or the radiative momentum luminosity $L_{Edd}/c$
exceeds $GMM_g/r^2$, i.e.  $\sigma^4/Gf_b$. Now $\dot M_w \propto L_{Edd}$
and $L_{Edd}=4\pi GcM_{bh}/\kappa.$ In contrast, for a supernova-driven wind:
$\dot M_w=\dot M_\ast E_{SN}(m_{SN}\sigma v_c)^{-1}.$ Assume now that
outflows lead to saturation of the star formation rate by exhausting the cold
gas supply.  I infer that $M_{bh}=\frac{\kappa\sigma^4}{4\pi G^2}.$  The cooling criterion for star formation efficiency
guarantees that this relation must saturate for black hole masses of around
$10^8\rm M_\odot$ if the relevant dynamical time-scale is gravitational   (corresponding to a spheroid mass of $\sim 10^{11}\rm M_\odot$), but
the reduced time-scale of AGN feedback increases the saturation limit to
$10^9-10^{10}\rm M_\odot.$

 If this is correct, the ULIG/ULIRG phenomenon involves both spheroid
formation and SMBH growth associated with the gas-rich proto-spheroid
phase.  The superwinds are AGN momentum-driven and are self-limiting,
with the rate of mass ejected inevitably being of order the star
formation rate. The SMBH-triggered associated outflows generate the
$M_{SMBH}\approx 10^6\sigma_7^4 \rm M_\odot$ relation, where $\sigma_7$
denotes the spheroid velocity dispersion in units of 100 km/s.
This is
in fact the observed correlation between $M_{SMBH}$ and $\sigma_g$ in both
slope and normalisation, naturally  cutting off above $10^9-10^{10} \rm M_\odot.$ 

Supernova-triggered galactic outflows are prevalent until
$\sigma_{gas}\approx 
100\rm km \, s^{-1}; $ at larger gas turbulence velocities, black hole
outflow-initiated outflows must dominate. Self-regulation of jet
outflow (positive) and star formation/SN (negative) feedback means that 
$$\dot M_w 
\sim
\dot M_\ast  \sim L_{Edd}/{\sigma v_w }\propto \sigma^3 v_w^{-1}
\simpropto v_w^2.$$
The predicted star formation rate is 
$$\dot M_\ast\approx \dot M_w (m_{SN}v_c\sigma/E_{SN})
\approx L_{Edd}(m_{SN}v_c/E_{SN}v_w).$$ 
The star formation luminosity is predicted 
to be of order $L_{stellar} \approx\dot M_\ast \epsilon_{nuc}f_{core},$
where $f_{core}$ is the mass in nuclear-burning stellar cores,
and hence  $$L_{stellar}/ L_{Edd}\approx \frac{\epsilon_{nuc}f_{core}E_{SN}}{m_{SN}v_cv_w\sigma}.$$
These represent predictions for ultraluminous star-forming galaxies at high
redshift that should eventually  be
 verifiable: 
the star formation rate  is proportional roughly 
to the square of the wind velocity
($\dot M_\ast \simpropto v_w^2$) and also 
 to the square root of the quasar luminosity  
 ($\dot M_\ast\propto  L_{Edd}^{1/2}$).

\section{ Observing cold dark matter: where next?}

There is a motivated dark matter candidate, the lightest stable SUSY
particle  under R parity conservation, or WIMP.  As yet, direct
detection experiments have not found any unambiguous evidence for its
existence. The Milky Way halo provides a laboratory par excellence for
indirect WIMP searches via annihilations into high energy particles
and photons.

The relic WIMP freezes out at $n_\chi<\sigma_{ann}v>t_H\simlt
1,$ corresponding to a temperature $T\simlt m_\chi/20k.$ The resulting CDM density is
$\Omega_\chi \sim \sigma_{weak}/\sigma_{ann}.$ Halo annihilations of the LSSP
occur into $\gamma$ and $\nu,$ as well as $\bar p, p$ and $e^+, e^-$
pairs. In fact, halo detectability may require clumpiness $\langle n^2
\rangle/\langle n \rangle^2\sim 100.$ SUSY modelling of parameter space
supplies the relation between $\sigma_{ann}$ and $m_\chi.$ There is an
uncertainty of some 2 orders of magnitude in the annihilation cross-section
at specified WIMP mass.  The WIMP mass most likely  lies in the range 0.1-10 TeV, and
annihilations provide possible high energy signatures via indirect detection for astronomy
experiments.  The only claimed evidence for direct detection 
relies on annual modulation in the DAMA NaI scintillation experiment,
which is
marginally viable for a spin-independent annihilation cross-section
and a low WIMP mass $(\sim 1-10 \rm GeV)$ \cite{ake05}.  
The uncertainties are large however, and improved data
is urgently needed to assess these issues.

One can envisage progress on a variety of fronts.  In particle theory,
one can readily imagine more than one DM candidate.  Why not have 2
stable dark matter particles, one light, one heavy, as motivated by
$N=2$ SUSY?  If one took the light dark matter and any of the possible
heavy dark matter detections seriously, one could have a situation in
which the light (a few MeV) spin-0 particle is subdominant but a $\sim
0.1-100 \, \rm TeV$ neutralino is the dominant relic \cite{fay04}.

Because a neutralino of mass $\simgt 1$ TeV is beyond the range of the
 LHC or even the ILC, astrophysical searches for DM merit serious
 consideration and modest funding.  In direct detection, one might
 eventually hope to see a modulated signal, due to the effect of the
 Earth's motion through directed streams of CDM \cite{fre05}. The
 streams are generic to tidal disruption of dark matter clumps.  As
 for indirect detection, the prospects are exciting, because of the
 many complementary searches that are being launched.  Evidence of
 neutralino annihilations may come from searches for $\gamma, \nu,
 e^+$ and $\bar p$ signatures. Experiments under development include
 HESS2, MAGIC, VERITAS, GLAST ($\gamma$-rays), ICECUBE, ANTARES,
 KM3NET ($\nu$), and PAMELA and AMS ($ e^+, \bar p).$ Targets include
 the Galactic Centre, the halo and even the sun, where neutralino
 annihilations in the solar core yield a potentially observable high
 energy neutrino flux \cite{sil86}.

 Refined numerical simulations
will soon explore the impact of supernova and SMBH-driven outflows and bar
evolution on the distribution and especially the concentration of
CDM. A better understanding of intermediate mass black holes 
 as well as the SMBH in the Galactic
Centre could eventually provide ``smoking guns" where spikes of CDM
were retained: the enhanced neutralino annihilations measure CDM where
galaxy formation began, 12 Gyr ago. Fundamental physics could be
probed: for example a higher dimensional signature, Kaluza-Klein dark
matter, would have a spectral signature and branchings that are
distinct from those of neutralinos.  The prospect of multi-TeV dark
matter is another tantalising probe. This provides a challenge for
SUSY but is possibly a natural and fundamental scale for any stable
relics surviving from n=3 extra dimensions.

\section{ Summary}

Galaxy formation is still poorly understood despite its apparent
successes. There is no fundamental theory of star formation. One can
adopt various empirical parameters and functions, incorporate
plausible assumptions and prescriptions and add new ingredients until
satisfactory explanations are obtained of any specified observations.
Beautiful images are often simulated at such vast cost in computer
time that it is impossible to test the robustness of the favoured
location in multidimensional parameter space.
 
Dark matter searches are not in a much heathier state. They rely on
plausible assumptions about the dark matter candidates
and on the theory of gravity. There is a
vast parameter space that admits undetectable particles, such as the
gravitino.  One has to hope that the likely culprit has electromagnetic 
couplings.

This is the down side. Bayesians would abandon hope at this juncture,
and argue that more science return per dollar will come, for
example, by sending men to Mars.  Yet to conclude on a more positive
note, there is every prospect that potential advances in  supercomputers, with
virtually no limit to the size of future simulations, will allow us to
reproduce our local universe in detail, thereby providing a firmer
basis for extrapolation to the remote past. And this extrapolation
could be largely phenomenological, driven by the data flow from ever
larger and more powerful telescopes that peer further into the
universe and hence into our past.

Likewise, the forthcoming LHC and the eventual construction of the ILC
will pose tighter constraints on the underlying particle physics that
provides the infrastructure for speculations about dark matter.  With
any luck, supersymmetry will be discovered, thereby setting dark
matter candidates on a far firmer footing. And the complementary
experiments in direct and in indirect detection should, within a decade,
probe all of the allowable SUSY parameter space.

This is an exciting moment in cosmology. We are at the threshold of
 confirming a standard model, which seems boring and even ugly. Yet
 the the prospect beckons of finding new physics in the unexpected
 deviations from the model. A convergence of particle physics and
 astronomy, in experiment and in theory, will inevitably lead us onto 
uncharted territory. There can be no greater challenge 
than in deciphering what awaits us.

  I thank my collaborators for discussions and exchanges on many of
the issues covered here.  

\end{document}